\documentclass[ reprint, aps, prl, two column, showkeys, groupedaddress]{revtex4-1}
\usepackage{graphicx,epsfig}
\usepackage{amssymb}
\usepackage{amsmath}
\usepackage{bm}
\usepackage{textcomp}
\usepackage{soul,xcolor}

\usepackage{color}

\begin{document}
\setcounter{page}{1}
\setstcolor{red}

\title[]{Design rules for modulation-doped AlAs quantum wells}
\author{Yoon Jang \surname{Chung}}
\author{K. W. \surname{Baldwin}}
\author{K. W. \surname{West}}
\author{D. \surname{Kamburov}}
\author{M. \surname{Shayegan}}
\author{L. N. \surname{Pfeiffer}}
\affiliation{Department of Electrical Engineering, Princeton University, Princeton, NJ 08544, USA  }
\date{\today}

\begin{abstract}

Thanks to their multi-valley, anisotropic, energy band structure, two-dimensional electron systems (2DESs) in modulation-doped AlAs quantum wells (QWs) provide a unique platform to investigate electron interaction physics and ballistic transport. Indeed, a plethora of phenomena unseen in other 2DESs have been observed over the past decade. However, a foundation for sample design is still lacking for AlAs 2DESs, limiting the means to achieve optimal quality samples. Here we present a systematic study on the fabrication of modulation-doped AlAs and GaAs QWs over a wide range of Al$_x$Ga$_{1-x}$As barrier alloy compositions. Our data indicate clear similarities in modulation doping mechanisms for AlAs and GaAs, and provide guidelines for the fabrication of very high quality AlAs 2DESs. We highlight the unprecedented quality of the fabricated AlAs samples by presenting the magnetotransport data for low density ($\simeq$ 1$\times$ 10$^{11}$ cm$^{-2}$) AlAs 2DESs that exhibit high-order fractional quantum Hall signatures.
\end{abstract}

\maketitle
Clean two-dimensional electron systems (2DESs) which exhibit the fractional quantum Hall effect are ideal systems to study electron-electron interaction phenomena and many-body ground-states. Along with the classic example of modulation-doped GaAs \cite{Tsui.PRL, Pan.PRL.2002}, recent studies have revealed that we can add AlAs \cite{Lay.APL,Etienne.Science,Ettiene.APL,Gunawan.Ballistic,ValleySkyrmions,ValleySusceptibility,Gunawan.PRB2006,Shayegan.Review,Bishop,Grayson.APL,Gokmen.NatPhys,Padmanabhan.PRL,Wegs1.PRB}, Si \cite{Lai.PRL.2004,Kott.PRB.2014}, Ge \cite{Shi.PRB.2015},  ZnO \cite{oxide}, and graphene \cite{Dean} to the list of materials in which high-order fractional states have been observed. The AlAs system is particularly exciting. First, its lattice constant closely matches that of GaAs, therefore allowing the growth of very high quality, single-crystal, AlAs epitaxial layers on GaAs substrates. Second, as shown in Fig. 1, AlAs distinguishes itself from GaAs in where its conduction-band electrons are in the first Brillouin zone. 

In bulk AlAs electrons occupy multiple conduction-band minima (valleys) with anisotropic energy vs wavevector dispersions. When electrons are confined to an AlAs quantum well (QW), by varying the well-width and in-plane strain, one can make the 2D electrons occupy the valleys with different (in-plane) anisotropy, effective mass, and effective Land\'{e} $g$-factor \cite{Shayegan.Review}. These different parameters, and the flexibility to control the valley occupation, render the AlAs 2DES a unique system for probing exotic many-body as well as ballistic transport phenomena. Recent studies in AlAs 2DESs have indeed  led to the observation of integer and fractional quantum Hall ferromagnetism \cite{Etienne.Science,Padmanabhan.PRL}, valley skyrmion formation \cite{ValleySkyrmions}, and interaction-enhanced valley susceptibility for electrons \cite{ValleySusceptibility,Shayegan.Review} and composite fermions \cite{Bishop}; it was also reported recently that the transport anisotropy of electrons is transferred to the composite fermions in AlAs QWs \cite{Gokmen.NatPhys}. The AlAs 2DES is a also a prime candidate for "valleytronic" devices \cite{Rycerz.Nat.Phys.2007}, and it was the first system where ballistic electron transport in different valleys was demonstrated \cite{Gunawan.Ballistic,Gunawan.PRB2006}.

\begin{figure*} [t]
  \begin{center}
    \psfig{file=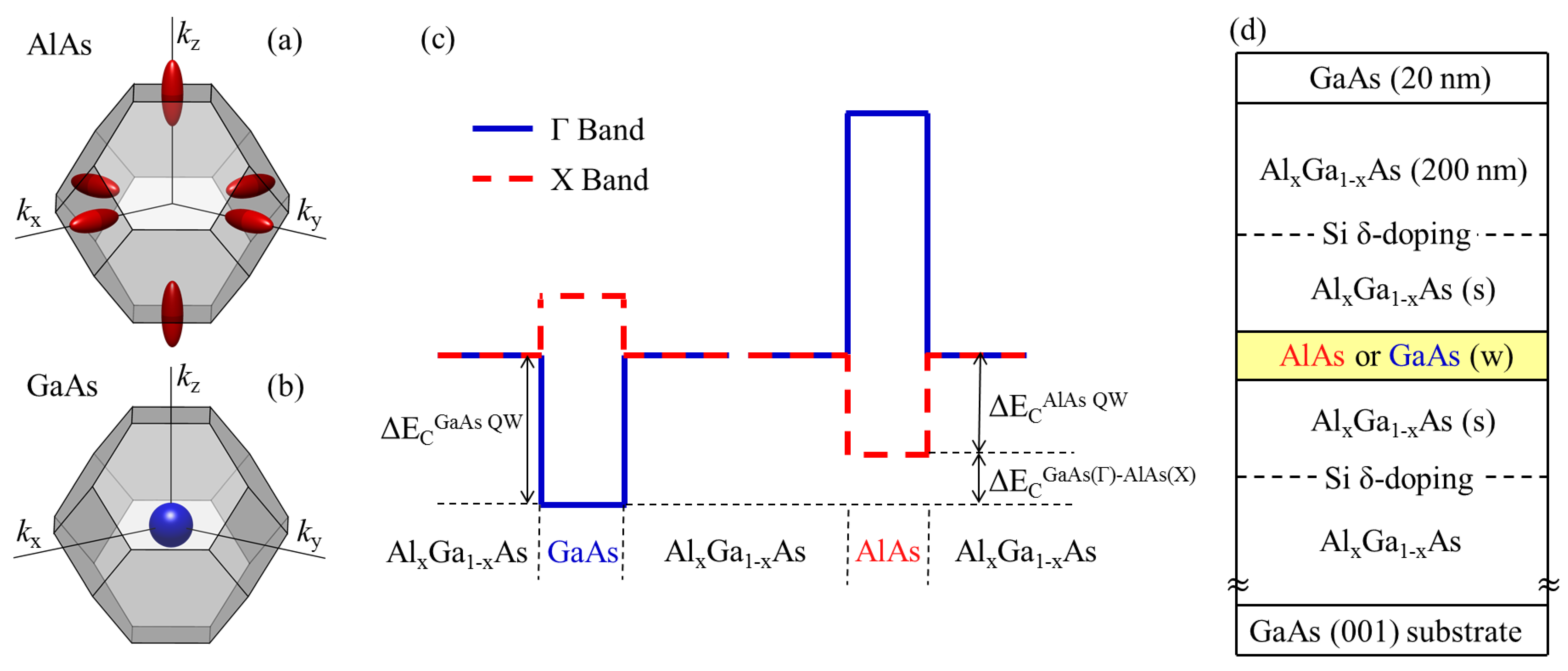, width=0.95\textwidth }
  \end{center}
  \caption{\label{fig1} The first Brillouin zones and electron Fermi surfaces of bulk (a) AlAs and (b) GaAs. (c) Comparison of the conduction band diagrams near GaAs and AlAs quantum wells. (d) The sample structure implemented in this work, with $w$ and $s$ denoting well width and spacer thickness, respectively}
\end{figure*}

Despite the abundance of literature concerning the rich physics of 2DESs in AlAs QWs, there are fundamental unanswered questions about modulation-doping in these systems. For example, over an extended period of time, many studies on AlAs QWs have utilized Al$_x$Ga$_{1-x}$As barrier alloy fractions in the vicinity of $x\simeq0.40$ \cite{Lay.APL,Etienne.Science,Ettiene.APL,Gunawan.Ballistic,ValleySkyrmions,ValleySusceptibility,Gunawan.PRB2006,Shayegan.Review,Bishop,Grayson.APL,Gokmen.NatPhys,Padmanabhan.PRL,Wegs1.PRB}. This choice is based on the fact that at this $x$ the minima of the $\Gamma$- and X-bands are known to cross, hence providing the maximum conduction band offset for populating the AlAs QW. However, as is well known for the case of GaAs QWs, maximum conduction band offset does not necessarily relate to the best sample quality because of factors such as interface quality or background impurities in the barrier \cite{MashMob,Mobility1}. As shown in Fig. 1(c), since the barrier material flanking an AlAs QW is similar to what flanks a GaAs QW except that the band minimum is the X-band rather than the $\Gamma$-band, we could expect similar behavior for AlAs QWs. However, because there have not been many studies on barrier alloy fractions other than $x\simeq0.40$, it is difficult to assess these possibilities. 

Here we provide guidelines to grow modulation-doped AlAs QWs, flanked by Al$_x$Ga$_{1-x}$As barriers with $0.20\leq x \leq 0.80$. By deducing the relevant energy levels from electron density measurements, we find that the modulation-doping characteristics of AlAs and GaAs QWs are essentially identical. Our data show that this is true over the entire range of $x$, where three different situations can occur for the conduction band alignment of the two types of QWs considering both the $\Gamma$- and X-band, as shown in Fig. 2. Because modulation-doping is a thermal equilibrium process, no fundamental distinction is observed when comparing cases that involve both the X- and $\Gamma$-bands (Figs. 2(a) and (f)) with the single X-band ($\Gamma$-band) processes in Figs. 2(b) and (c) (Figs. 2(d) and (e)). We highlight this fact by demonstrating high quality modulation-doped AlAs QWs with $x=0.33$.

\begin{figure}[t]
  \begin{center}
    \psfig{file=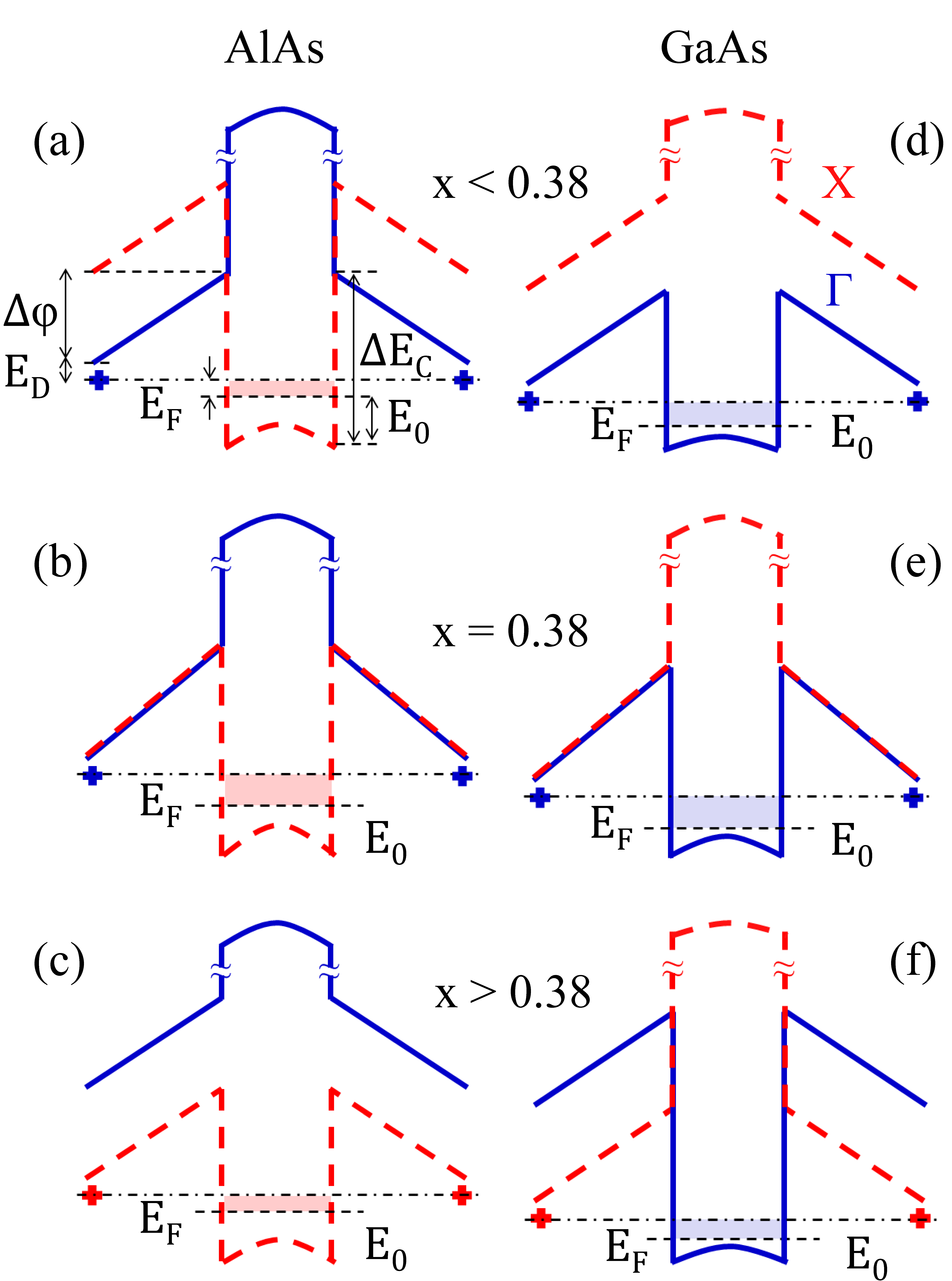, width=0.45\textwidth }
  \end{center}
  \caption{\label{fig2} Schematic diagrams of the conduction band in the vicinity of (a)-(c) AlAs and (d)-(f) GaAs quantum wells for barrier alloy fractions $x<0.38$, $x=0.38$, and $x>0.38$. The dashed red and solid blue lines represent the X-band and $\Gamma$-band edges, respectively. }
\end{figure} 

\begin{figure*}[t]
  \begin{center}
    \psfig{ file=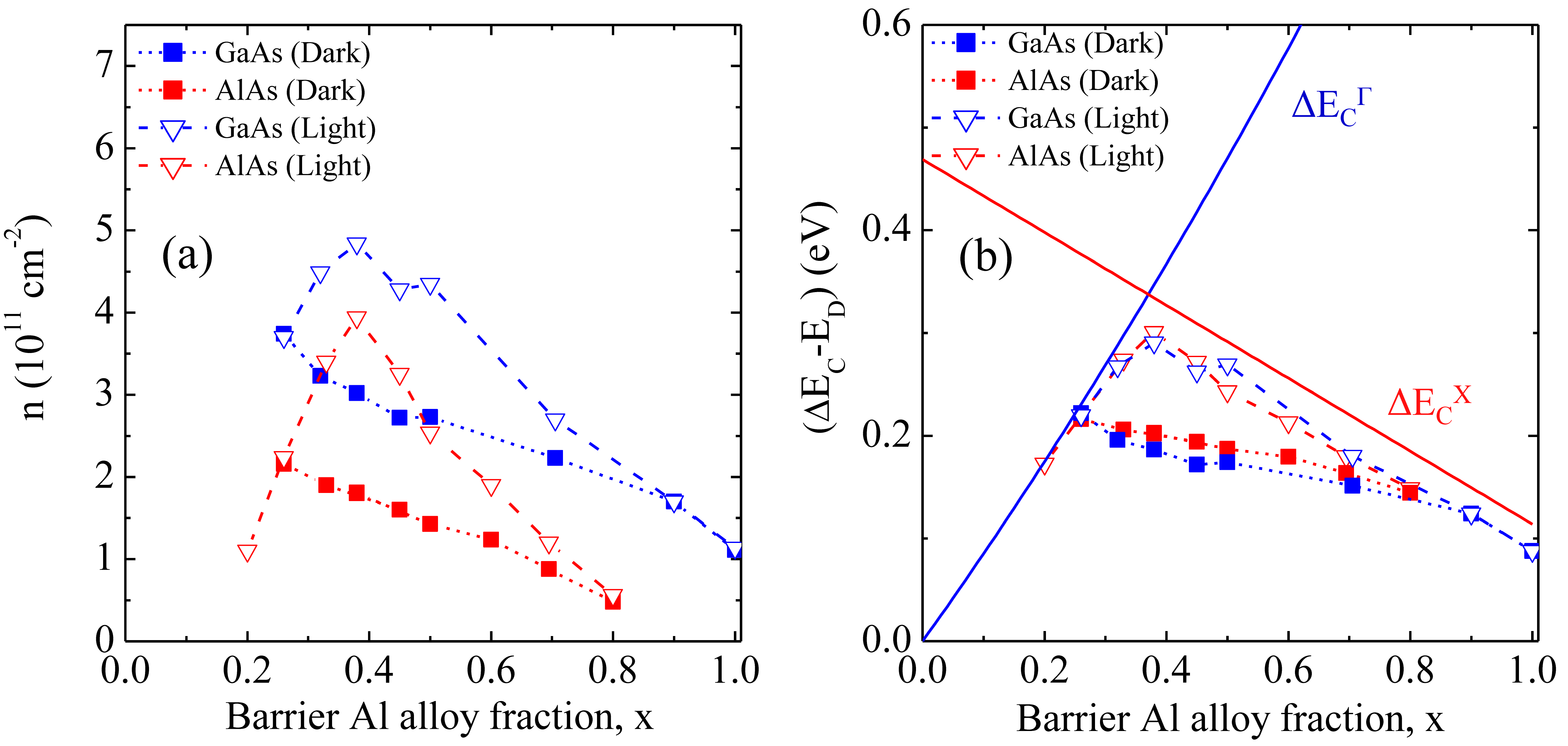, width=0.95\textwidth }
  \end{center}
  \caption{\label{fig3} (a) Measured electron densities for the AlAs and GaAs QWs as a function of $x$. The squares represent densities measured in the dark while the triangles show data points after light exposure; the lines are guides to the eye. The well-width and spacer thicknesses are 11 and 59 nm for the AlAs and 20 and 70 nm for the GaAs samples. (b) Values of ($\Delta$$E_C-E_D$) deduced from the experimental data points of Fig. 3(a) for the AlAs and GaAs QW samples (see text). The solid blue and red lines show our estimates for the $\Gamma$- and X-band edge energies relative to the $\Gamma$-band edge of GaAs.}
\end{figure*}

\begin{figure} [t]
  \begin{center}
    \psfig{file=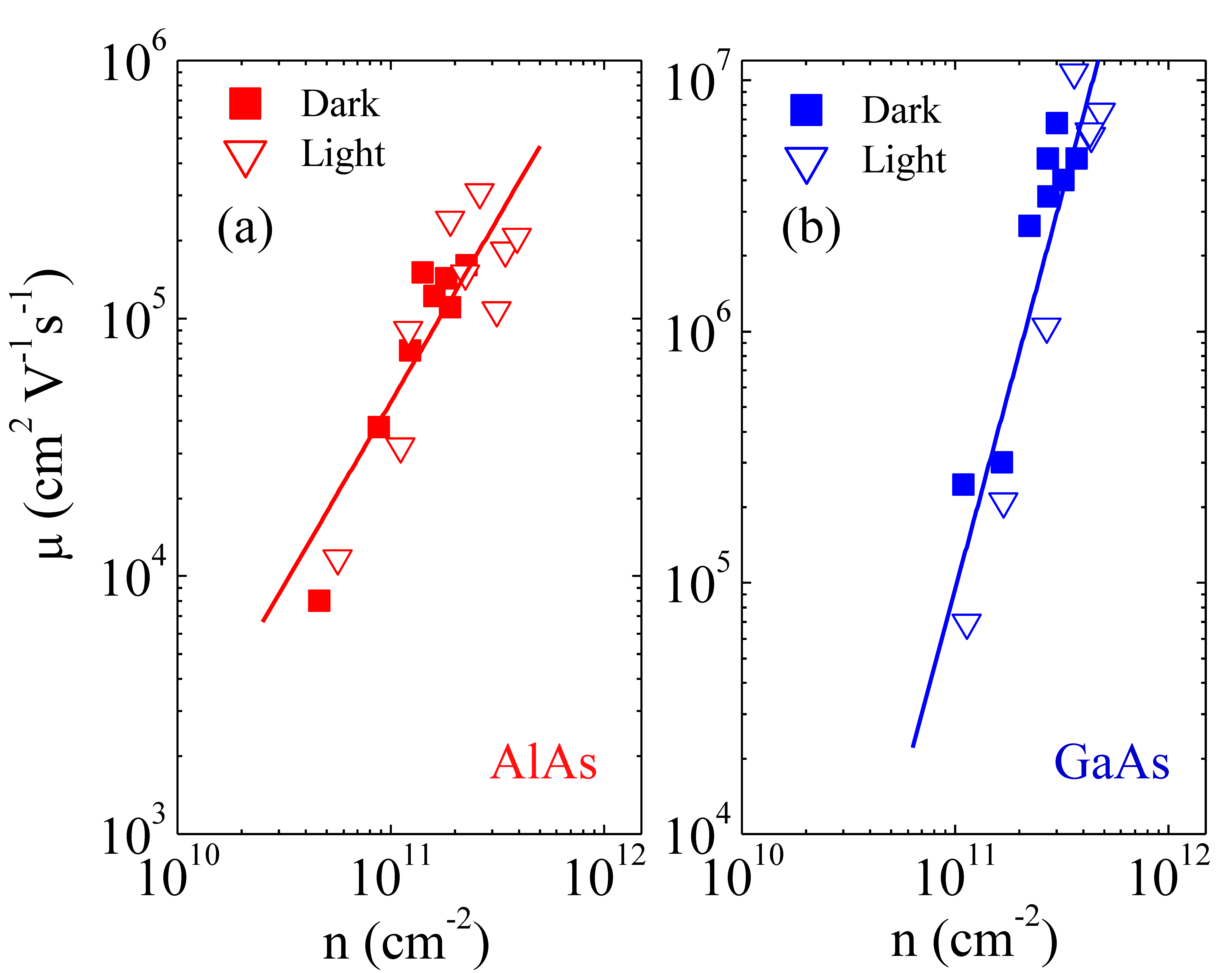, width=0.45\textwidth }
  \end{center}
  \caption{\label{fig4} Measured mobility values for the (a) AlAs and (b) GaAs QWs in our study. The power law fits correspond to a relation of $\mu\propto n^{1.4}$ for the AlAs QWs and $\mu\propto n^{3}$ for the GaAs QWs.}
\end{figure}

\begin{figure*}[t]
  \begin{center}
    \psfig{file=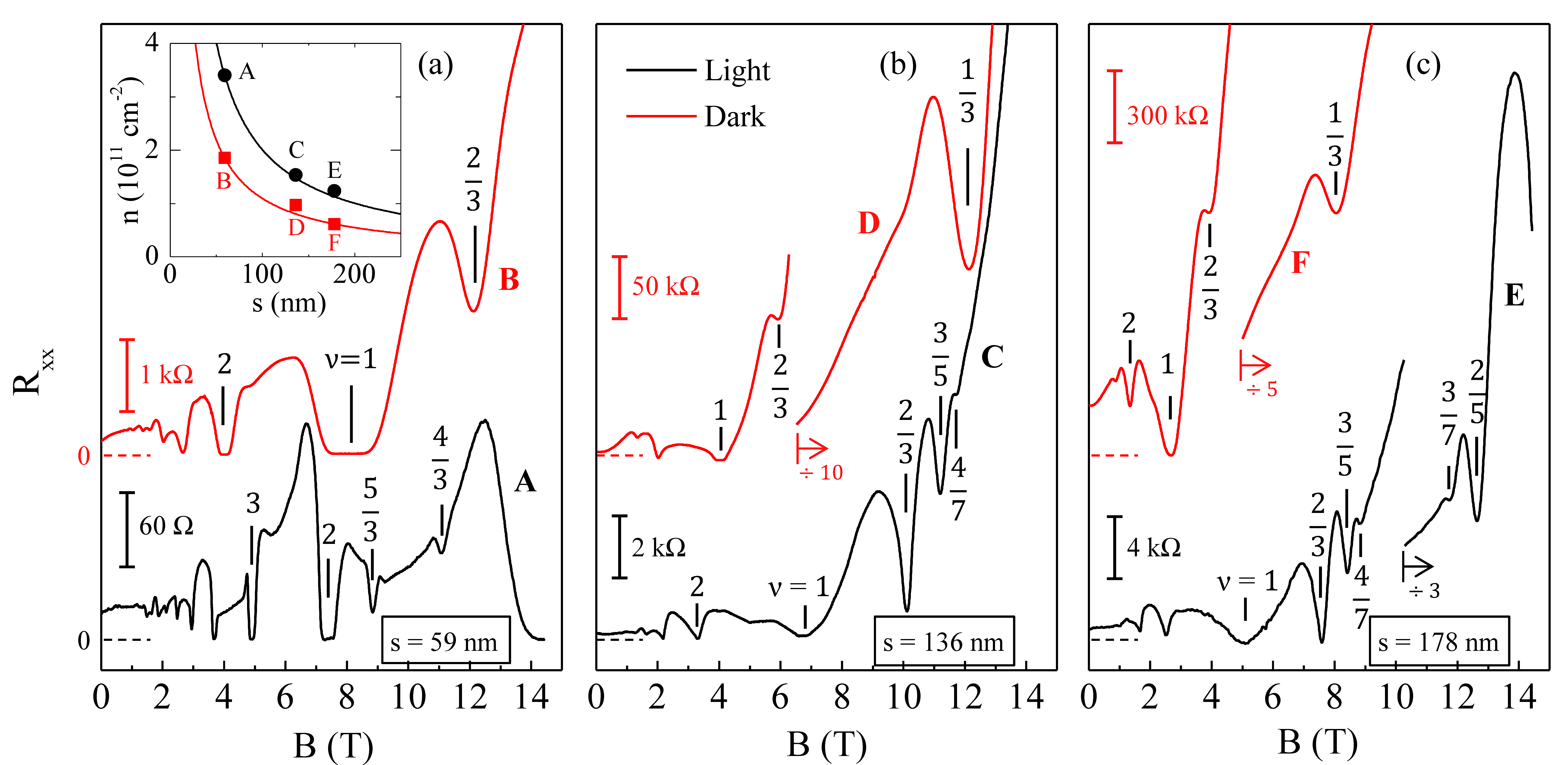, width=0.95\textwidth }
  \end{center}
  \caption{\label{fig5} (a)-(c) Representative magnetotransport ($R_{xx}$) traces measured at 0.3 K in the dark and after illumination for AlAs QW structures with spacer thicknesses of 59, 136, and 178 nm. The inset in (a) shows the electron densityl vs spacer thickness, with solid lines representing the expected values for $n\propto s^{-1}$ [see Eq. (1)]. The scales for the resistance axes are given for each trace with dashed lines denoting zero. The Landau level filling factors ($\nu$) for some of the integer and fractional quantum Hall states are marked  in each trace. The indices of {\bf A}-{\bf F} in the inset to (a) are given to mark the corresponding electron densities of each trace; these densities are {\bf A}: 3.4, {\bf B}: 1.9, {\bf C}: 1.5, {\bf D}: 9.7, {\bf E}: 1.2, {\bf F}: 0.61, all in the units of $10^{11}$ cm$^{-2}$.}
\end{figure*}

For our study, AlAs or GaAs QWs, flanked by Al$_x$Ga$_{1-x}$As barriers with $\delta$-Si doping, were grown by molecular beam epitaxy on (001) GaAs substrates [see Fig. 1(d)]. We use a Si doping concentration ranging from $\simeq$ $3\times 10^{11}$ cm$^{-2}$ to $1\times 10^{12}$ cm$^{-2}$ for the substrate side, and $\sim$ 1.5 to 2 times this value on the surface side. The lower limit is implemented to prevent parallel conduction in the $x\leq0.26$ AlAs QWs. The growth temperature was measured by a factory calibrated optical pyrometer (Ircon Modline 7V-1205, emissivity set to 0.63) and was fixed to be 645\textdegree C for all samples at all times except for when $\delta$-doping the 2 nm Al$_x$Ga$_{1-x}$As layer beneath the lower spacer of the QWs, where the temperature was lowered to 480\textdegree C to prevent surface segregation of the Si \cite{A,B,C,D}. The $x$ ranged from 0.20 to 0.80 for the AlAs QWs and 0.26 to 1.0 for the GaAs QWs. We examined reflection high energy electron diffraction patterns prior the growth of each sample to determine compositions and growth rates. Well width ($w$) and spacer thickness ($s$) were fixed at 11 and 59 nm for the AlAs QWs and 20 and 70 nm for the GaAs QWs. For measurements we used a low frequency lock-in technique and a pumped $^{3}$He cryostat with a base temperature of 0.3 K. Magnetoresistance data were taken by sweeping a superconducting magnet from 0 to 14.5 T in the dark and after illuminating the sample with a red light emitting diode at $\sim$10 K.  

Before presenting the experimental data, we briefly describe the valley occupation and parameters for our AlAs 2DESs. In AlAs QWs with $w\gtrsim5$ nm, biaxial compression from epitaxial growth on GaAs substrates raises the ground-state energy of the valley with its major axis along the growth direction, causing the other two (in-plane) valleys to be occupied \cite{Shayegan.Review}. Our AlAs QWs have a well width in this regime and thus have in-plane effective mass values of $m{_l}^*=1.1m_e$ and $m{_t}^*=0.20m_e$, with a geometric mean of $m^*=\sqrt{m{_l}^*m{_t}^*}=0.45m_e$, and an out-of-plane mass of $m{_t}^*=0.20m_e$ ($m_e$ is the free electron mass).

Figure 3 (a) shows the density of electrons ($n$) for our AlAs and GaAs QWs as a function of $x$. All electron concentration values were evaluated from the quantum Hall features in the magnetotransport data. Although there is an offset between the density profiles for the two QWs, it is clear that the variations in $n$ for GaAs and AlAs QWs have a similar trend with $x$ for measurements both taken in the dark and after light illumination. As we elaborate below, this suggests a common mechanism for the modulation-doping of the two QWs.

It is important to note here that at higher barrier alloy fractions ($x \geq 0.38$), an annealing technique \cite{Grayson.APL} is required to achieve saturated carrier concentrations after illumination. Different annealing conditions are needed for saturation for different $x$, with $x=0.38$ having the longest time-constant of the order of 1 hour at $\sim$40 K. For $x\leq0.33$, the extra annealing step was unnecessary, likely because the time-constant is short enough so that the process is completed during the $\sim$30 minutes it takes to cool the illuminated sample from 10 K to 0.3 K in our system \cite{NegPPCGaAs.APL}. This behavior is observed for both GaAs and AlAs wells, corroborating our conclusion that the modulation-dopings into these wells share a universal mechanism.

Using the textbook model for modulation-doped heterostructures \cite{Davies.Book}, we can relate our measured $n$ with the energy levels of the QW:
 \begin{align}
  \Delta E_C=E_0+E_F+E_D+\Delta\varphi
   \label{eqn:1}
 \end{align}
\noindent where $E_0$ is the ground-state energy measured relative to the conduction band edge of the QW, $E_F$ is the Fermi level measured with respect to $E_0$, $E_D$ is the donor level energy defined relative to the conduction band edge of the barrier, and $\Delta\varphi\equiv{nse^2}/{\epsilon_{b}}$ [see Fig. 2(a)]. Here $\epsilon_{b}$ is the barrier dielectric constant and $e$ is the electron charge. Using values of $n$ and $s$ we can determine $\Delta\varphi$ and $E_F=n\pi\hbar^{2}/g_{v}m^{*}$, where $\hbar$ is the Planck constant and $m^{*}$ is the effective mass in the QW ($m^{*}=0.067m_e$ for GaAs and $m^{*}=0.45m_e$ for AlAs); $g_{v}$ is the valley degeneracy ($g_{v}=1$ for GaAs and $g_{v}=2$ for AlAs). From the simple case of an infinite potential well, we can also get a rough estimate for $E_0$, which is $\simeq$ 15 meV for the AlAs QWs ($m{_t}^*=0.2m_e$, $w=11$ nm) and $\simeq$ 14 meV for the GaAs QWs ($m^*=0.067m_e$, $w=20$ nm). Considering, as an example, the case of the GaAs QW with $x=0.33$ and $n=4.5\times10^{11}$ cm$^{-2}$ after illumination, we deduce $\Delta\varphi\simeq238$ meV, and $E_F\simeq16$ meV. Since $E_0$ and $E_F$ are both much smaller than $\Delta\varphi$, we conclude from Eq. (1) that $\Delta\varphi\simeq(\Delta E_C-E_D)$. Implementing a self-consistent Schr\"{o}dinger-Poisson solver corrects $E_0$ of the order of $\simeq5-10$ meV. More precise calculations require an exact knowledge of $\Delta$$E_C$ for all alloy fractions, but this would not alter the relation $\Delta\varphi\simeq(\Delta E_C-E_D)$ which is the crucial factor in understanding the design rules in this study.

The symbols in Fig. 3(b) show the values of ($\Delta$$E_C-E_D$), with respect to the $\Gamma$-band edge of GaAs, deduced from the density data points in Fig. 3(a) and using Eq. (1) assuming $E_0$ values of an infinite potential well. To account for the fact that the conduction band minima of AlAs QWs are not aligned with GaAs QWs, we take the offset between GaAs($\Gamma$) and AlAs(X) to be 114 meV \cite{130meV} and add this constant value to all the ($\Delta$$E_C-E_D$) values for the AlAs QWs. Since the calculated $\Delta\varphi$ is $\simeq70$ meV for our AlAs/GaAs/AlAs (i.e., $x=1$) structure, and previous reports quote shallow donor energies in AlAs ranging from $\simeq30$ to 60 meV \cite{Chand.PRB,Adachi.JAP,Bulk.PRB}, the value of 114 meV we take from the literature is quite consistent with our results. It is seen in Fig. 3(b) that with the 114 meV offset there is excellent agreement between the ($\Delta$$E_C-E_D$) values for the AlAs and GaAs QWs over the entire range of $x$. From our data points measured after illumination, we can estimate the conduction band offset with respect to GaAs($\Gamma$) for Al$_x$Ga$_{1-x}$As in modulation-doped structures, drawn as the solid blue and red lines for the $\Gamma$- and X-bands in Fig. 3(b). For the $\Gamma$-band, the $x<0.38$ data coincide very well with the reported literature values of the conduction band offset $\Delta$$E_{C}^{\Gamma}$ \cite{Pav.JAP,Batey.JAP} assuming a hydrogenic donor level. For the X-band we draw a line that goes through the reported GaAs(X)-GaAs($\Gamma$) offset value of $\simeq470$ meV \cite{Pav.JAP} and our expected AlAs(X)-GaAs($\Gamma$) offset of 114 meV. We find there is reasonable agreement with the data in Fig. 3(b), including the $\Gamma$-X band crossing point in Al$_x$Ga$_{1-x}$As at $x=0.38$. The deep donor levels measured from the data in the dark agree well with previous reports on the DX effect in Al$_x$Ga$_{1-x}$As \cite{Chand.PRB,Chadi.PRB}, showing a maximum effective barrier near $x=0.26$ and monotonic decrease when $x>0.26$.

We also comment on the mobility values measured for our samples. Figures 4(a) and (b) show the mobility values as a function of carrier concentration for the AlAs and GaAs samples, respectively. Note that for the AlAs samples at any given density, the measured mobility is higher than in previous studies \cite{Grayson.APL,Ettiene.APL}, attesting to the high quality of the samples used in our study. This is particularly impressive considering that, in contrast to the samples in \cite{Ettiene.APL}, our samples are doped from both sides and have smaller spacer thicknesses. The power law fit for the relation between density and mobility yield $\mu\propto n^{1.4}$ for the AlAs QWs and $\mu\propto n^3$ for the GaAs QWs. We postulate that the notable deviation from the well known $\mu\propto n^{1.5}$ for the GaAs samples is due to significant contributions from the barrier in the two lowest density samples, where $x=1.0$ and 0.9. Indeed if we perform a fit while omitting the data from these two samples, we achieve a power law of $\mu\propto n^{1.6}$. These results suggest that barrier quality is also an important factor to consider in sample optimization as mentioned earlier in the introduction.

To evaluate the potential of high quality AlAs samples with $x<0.38$, we grew a set of AlAs QWs with $x=0.33$ and varying spacer thicknesses. Figures 5(a)-(c) show longitudinal magnetoresistance ($R_{xx}$) data for the $x=0.33$ AlAs wells with spacer thicknesses of 59, 136, and 178 nm, respectively. The dependence of $n$ on spacer thickness is plotted in Fig. 4(a) inset, which clearly shows that it is governed by the $n\propto s^{-1}$ relation expected of modulation-doped structures. The indices {\bf A}-{\bf F} in Figs. 5(a)-(c) and the inset mark the corresponding densities of each trace.

Figures 5(a)-(c) demonstrate the high quality of the fabricated samples, with clear indications of the 2/3 and 1/3 fractional quantum Hall states (FQHSs) even at the low density of $6.1\times10^{10}$ cm$^{-2}$ ({\bf F}) for the 178-nm-spacer sample [Fig. 5(c)]. After light exposure $n$ for this sample increases to $1.2\times10^{11}$ cm$^{-2}$ ({\bf E}), and the measured trace shows excellent quality with clear $R_{xx}$ minima at filling factors $\nu=2/3$, 3/5, 4/7, 3/7, and 2/5. The $s=59$ nm sample which has a higher density of $3.4\times10^{11}$ cm$^{-2}$ ({\bf A}) after light also shows FQHSs at $\nu=5/3$ and 4/3. We emphasize that all of these samples were fabricated with $x=0.33$, and were measured after a brief illumination of $\sim1$ minute (with a current of 6 mA in the light emitting diode) at 10 K and a subsequent cool-down to 0.3 K after the light was turned off, with no additional procedures such as annealing \cite{Grayson.APL} or gating \cite{Ettiene.APL}. The results presented here suggest that when the conditions of the barrier are dominant in determining the quality of the AlAs 2DESs, we can resort to the conventional techniques used for GaAs to implement small $x$ barriers. For example, in GaAs samples with sufficiently large $s$, the intentional ionized impurities are far enough from the 2DES that the scattering term from the unintentional (background) impurities in the barrier becomes significant, and hence having a small $x$ barrier is crucial for a high quality due to the inherently more reactive nature of Al compared to Ga. If we extend this concept to AlAs QWs, it suggests that in low density AlAs 2DES where $s$ is large, we should grow AlAs QWs with small $x$ for optimal performance. This could also apply to narrow AlAs QWs, where the significant penetration of the electron wave function into the barrier again makes it beneficial to have a barrier with small $x$.

In conclusion, our measurements of the electron density in modulation-doped AlAs and GaAs QWs over a wide range of Al$_x$Ga$_{1-x}$As barrier alloy fractions reveal that their doping characteristics are essentially identical despite having different electron pocket distributions in the Brillouin zone. We highlight this by the observation of the $n\propto s^{-1}$ rule for $x=0.33$ AlAs wells with 59, 136, and 178 nm spacer thicknesses. Our fabricated AlAs QWs show high quality magnetotransport data with clear indications of FQHSs. The design rules we establish here for modulation-doped GaAs and AlAs QWs provide a foundation for application specific sample optimization, especially in the case of AlAs which was so far a relatively uncharted material compared to GaAs. 
 
\begin{acknowledgments}
We acknowledge support through the NSF (Grants DMR 1305691 and ECCS 1508925) for measurements, and the NSF (Grant MRSEC DMR 1420541) and the Gordon and Betty Moore Foundation (Grant GBMF4420) for sample fabrication and characterization.
 \end{acknowledgments}

\end{document}